\documentclass[twocolumn,pre,aps]{revtex4}
\usepackage{epsfig}
\begin{document}
\date{\today}

\title{Curved Tails in Polymerization-Based Bacterial Motility}
\author{Andrew D. Rutenberg}
\homepage{http://www.physics.dal.ca/~adr/profile.html}
\affiliation{Department of Physics, Dalhousie University, 
Halifax, NS, Canada B3H 3J5}
\author{Martin Grant}
\affiliation{Centre for the Physics of Materials, Physics Dept.,  
3600 rue University,
McGill University, Montr\'{e}al QC, Canada H3A 2T8}

\begin{abstract}
The curved actin ``comet-tail'' of the bacterium {\em Listeria monocytogenes} 
is a visually striking signature of actin polymerization-based motility. 
Similar actin tails are associated with {\em Shigella flexneri}, spotted-fever
{\em Rickettsiae}, the {\em Vaccinia} virus, and vesicles and microspheres
in related {\em in vitro} systems. We show that the torque required to produce 
the curvature in the tail can arise from randomly placed actin filaments 
pushing the bacterium or particle.  We find that the 
curvature magnitude determines the number of actively pushing 
filaments, independent of viscosity and of the molecular 
details of force generation.  The variation of the curvature with time 
can be used to infer the dynamics of actin filaments at the bacterial surface. 
\end{abstract}
\maketitle

\section*{Introduction}

The bacteria {\em L. monocytogenes}, {\em S. flexneri}, 
the spotted fever group of {\em Rickettsiae}, and the {\em Vaccinia} virus
are intracellular pathogens that move through the continual
polymerization of actin \cite{Theriot95,Dramsi98,Heinzen99,Cudmore95} in
distinctively curved ``comet-tails'' of actin filaments behind the 
motile particles.  While fascinating on its own, the 
actin comet-tail is functionally similar to the 
actin mesh in the lamellipodia of a locomoting eukaryotic cell, and the 
bacterial surface is analogous to the leading edge of the cell. 
Identification of the biochemical components involved has thus provided 
insight into the active regulation of actin polymerization by the cell
\cite{Machesky99a,Loisel99} --- an essential cellular process \cite{Alberts94}.

For polymerization-based motility, 
the force generated by actin polymerization at the moving object's surface 
drives the object forward against the viscous drag of the cytoplasm 
\cite{Mogilner96,Gerbal99}.  The necessary and sufficient bacterial
contribution to motility is a single surface protein that 
orchestrates cellular cytoplasmic proteins to locally promote the nucleation, 
elongation, and cross-linking of actin filaments. In {\em L. monocytogenes}, 
this process is driven by the bacterial protein ActA
\cite{Kocks92,Smith95,Domann92,Brundage93},
while {\em S. flexneri} expresses the 
protein IcsA for the same purpose \cite{Goldberg93,Goldberg95}.
Candidates for similar proteins have been proposed for
spotted-fever {\em Rickettsiae} \cite{Heinzen93,Heinzen99} and for the 
{\em Vaccinia} virus \cite{Frischknecht99}.
Simplified systems that have been developed 
for the study of polymerization-based motility include
{\em Escherichia coli} expressing IcsA on their surfaces 
\cite{Goldberg95,Kocks95} and microspheres coated with purified 
ActA \cite{Cameron99}.  Similar motility mechanisms appear to be at work in 
endosomal rocketing \cite{Merrifield99},  and in non-actin 
polymerization-based motility systems derived from nematode 
sperm \cite{Italiano96}.  Actin polymerization-based motility 
may even play an important role in vesicle trafficking within 
the cell \cite{Ma98,Rozelle2000}. 

As the bacterium or particle is driven forward, a curved comet-like tail of 
actin filaments remains behind. Photobleaching experiments in 
{\em L. monocytogenes} \cite{Theriot92}
and qualitative observation of {\em S. flexneri} \cite{Goldberg95} 
and of spotted-fever {\em Rickettsiae} \cite{Heinzen99}
demonstrate that the tail is stationary with respect to the surrounding
environment, probably due to steric or functional connections with the 
cellular cytoskeleton, so that the shape of the tail represents the 
path of the bacterium.  The curvature of this path varies from bacterium 
to bacterium, and changes over time for individual bacteria.  It is not 
known what determines the bacterial path, and hence the tail curvature, 
though no active control or chemotactic behavior has been proposed for 
these systems. 

Bacteria are functioning micromachines, but cannot be fully exploited 
without being fully understood.  The {\em L. monocytogenes} motility system 
is well enough understood biochemically that ActA coated microspheres 
\cite{Cameron99} should reconstitute polymerization-based motility in 
solutions of purified proteins \cite{Loisel99}, i.e. with total experimental 
control. However, we do not yet {\em quantitatively} understand the motility 
enough to be able to use the bacterial or microsphere motion as a probe of 
the bacterial or cellular conditions, or conversely to attempt to tailor those
conditions to affect the bacterial motion. 

In this paper, we propose that the curvature results from the random
location of actin filaments pushing against the bacterium. 
We show how the curvature of the bacterial path can be used to predict 
static and dynamic structure at the bacterial surface.  The average curvature
is determined by the number of active filaments pushing the 
bacterium.  Information about filament lifetime and surface diffusion rates 
may be obtained from curvature autocorrelations, since curvature 
depends on the location of active filaments with respect to the bacterial 
surface. If filaments are closely localized to specific bacterial surface 
proteins, then lifetimes and diffusivities of those proteins can be inferred. 
We focus on the most common experimental geometry of a thin 
quasi-two-dimensional system constrained between a glass slide and cover-slip, 
however we also discuss what would be expected for bulk geometries.  
In both cases, we discuss the apparent curvature appropriate to video 
microscopy.

\section*{Curved Trajectories}

\begin{figure}
\centerline{\epsfxsize = 3.25truein
\epsfbox{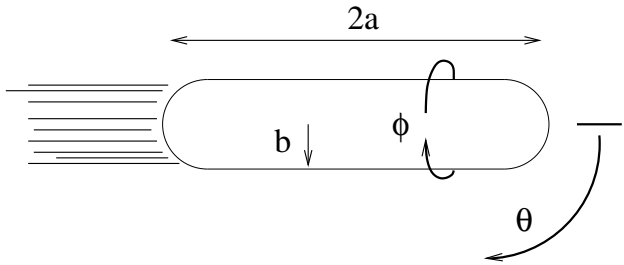}}
\caption{The polar angle $\theta$ and the
axial angle $\phi$ with respect to the capsule-shaped bacterium. 
The bacterium is taken to be curving within the plane of the page. 
The half-length and the radius of the bacterium are $a$ and $b$, respectively.
A schematic actin comet-tail is shown to the left, and the bacterium is 
pushed towards the right.  We approximate the bacterium as a prolate ellipsoid 
for purposes of drag calculations. 
\label{FIG:angles}}
\end{figure}

Curvature is defined as the rate of rotation of direction,
$K \equiv d\theta/ds$, or the rotation per unit path length, 
where $\theta$ is the polar angle in the current plane of motion 
(see Fig.~\ref{FIG:angles}) and $s$ is 
distance measured along the particle path.  The radius of curvature,
$R\equiv 1/|K|$, equals the radius of the circle that locally best fits the 
path. Either $K$ or $R$ {\em locally}
characterize the path, i.e. both may vary along the bacterial path. 

Curved trajectories imply a torque $N$ is acting on the bacterium 
to balance viscous drag proportional to 
the angular rate of rotation, $N = \dot{\theta} C_{turn}$,
where $C_{turn} = (32 \pi/3)\eta (a^4-b^4)/ [(2a^2-b^2)S-2a]$ 
is a rotational drag coefficient for turning in $\theta$,
and $\eta$ is the fluid viscosity, $a$ and $b$ are bacterial dimensions
(see Fig.~\protect\ref{FIG:angles}), and
$S = 2 (a^2-b^2)^{-1/2} \ln [(a+\sqrt(a^2-b^2))/b]$ \cite{Perrin34}. 
The viscous drag force $F$ due to linear motion at speed $v$
also carries a factor of the viscosity, where $F = v f_{lin}$ and 
$f_{lin} = 16 \pi \eta (a^2-b^2)/ [(2a^2-b^2)S-2a]$ \cite{Perrin34}.
Since the curvature of the path, 
$K = \dot{\theta}/v = (N/C_{turn})/(F/f_{lin})$, 
we can use the drag coefficients to obtain a remarkably simple result, 
\begin{equation}
\label{EQN:curvature}
		K = \frac{3}{2(a^2+b^2)} \frac{N}{F}, 
\end{equation}
independent of the viscosity.  This is fortunate since the 
effective viscosity of the cellular cytoplasm is strongly scale-dependent, 
ranging from $0.01$ poise ($10^{-3} Pa\;s$) for small loops on dye molecules 
\cite{Lubyphelps93}
to $2100$ poise ($210 Pa\;s$) for $1.3 \mu m$ diameter spheres \cite{Bausch99}. 
We find that the curvature directly probes the ratio of force 
and torque applied to the bacterium, with an easily determined 
geometrical prefactor.

The force, $F= f_0 n$, is proportional to the number of actively
pushing filaments $n$, while the force per filament 
$f_0$ depends on the specific details of the motility mechanism --- as seen
explicitly in thermal-ratchet models of polymerization-based motility 
\cite{Mogilner96}. A complementary coarse-grained 
elastic analysis of polymerization-based bacterial motion \cite{Gerbal99} 
exists, however it is 
not convenient for determining curvatures. We take $f_0$ as constant in 
time, which amounts to considering only times much greater than the mean-time 
between actin monomer addition or equivalently distances much greater than the 
monomer size $2.7nm$.  This is appropriate, since observed radii of curvature 
are larger than the bacterial scale (which is $1-2$ microns).

To calculate the torque $N$, we must consider the torque due to each filament.
These individual torques depend on exactly where the filament pushes on 
the bacterium.  We assume that the  $n$ actively pushing 
filaments are each randomly placed on the trailing end of the bacterium, 
so that each one will produce a random vectorial torque on the bacterium. 
The sum of many of these random torques will have a 
Gaussian distribution with zero mean. We can calculate
the root-mean-square (RMS) torque, $N_{rms}$, from the local filament density.
We take the growing filament barbed-ends as uniformly distributed over the
hemispherical cap at the end of the bacterium.  In cross-section, this 
leads to enhanced filament density at the edges of the tail, similar to that
seen in thin-section electron micrographs of {\em L. monocytogenes}
\cite{Tilney92}.  This distribution also follows naturally if the filament 
density follows a uniform ActA surface density.  In cross-sectional 
coordinates, where the cylindrical radius $r$ ranges from $0$ to $b$, the 
filament density distribution is 
\begin{equation}
\label{EQN:distrib}
		P_f(r,\phi) = \sigma b /\sqrt{b^2-r^2},
\end{equation}
where $P_f r dr d\phi$ is the average number of filaments
in the interval $(r,r+dr)$ and $(\phi,\phi+d\phi)$. Here, 
$\sigma = n/(2 \pi b^2)$ is the uniform surface filament density 
on the hemispherical end of the bacterium, and $b$ is the bacterial radius.  
The mean-square torque perpendicular to the direction of motion is easily found
\begin{equation}
\label{EQN:torque}
	\langle N^2 \rangle = f_0^2 \int_0^{2\pi} \int_0^b d\phi r dr
					[(r \sin \phi)^2 +(r \cos \phi)^2] 
					P_f(r,\phi),
\end{equation}
where $f_0 r \sin \phi$ and $f_0 r \cos \phi$ are the two components of the
torque, so that they add in quadrature.  This leads to an RMS torque 
$N_{rms} \equiv \sqrt{ \langle N^2 \rangle} = f_0 b^2 \sqrt{4 \pi \sigma/3}$.  
Taking the ratio $N_{rms}/F$ to calculate 
the RMS curvature from Eq.~\ref{EQN:curvature}, we have 
\begin{equation}
\label{EQN:rmsK}
		K_{rms} = \frac{\sqrt{3}}{(a^2+b^2) \sqrt{4\pi \sigma}}
				= \frac{b}{a^2+b^2} \sqrt{\frac{3}{2n}}
\end{equation}
for a hemispherical distribution of filaments.  Bacterial size 
and shape contribute to the curvature, as does the average distribution
of filaments Eq.~\ref{EQN:distrib}.  Other surface distributions of 
filament densities are also possible, and would affect the geometric 
prefactors in Eq.~\ref{EQN:rmsK} though not the functional dependence on 
the number of filaments $n$.  

Remarkably, the average force per filament 
$f_0$ does not appear in our expression for the curvature, so that our 
results appear {\em independent} of the details of the force generation 
mechanism.  However within polymerization ratchet models 
\cite{Mogilner96} thermal fluctuations of semi-flexible actin filaments
transverse to their length \cite{Kroy96} could generate transverse forces, 
which would lead to an increased torque and greater curvature than predicted a
bove in Eqn.~\ref{EQN:rmsK}.  This transverse contribution would depend on the 
biomechanics of the coupling between the actin filament and the bacterial 
surface, which would also depend on the bacterial shape. Unfortunately it 
also depends on the effective and anisotropic elastic constants of the actin 
filaments \cite{Kroy96}, which in turn sensitively depends on 
how actively pushing actin filaments are cross-linked into the 
bacterial tail and the cytoskeleton --- as can be seen by contrasting
the elastic constants given by \cite{Kroy96} and \cite{MacKintosh95}.
Our simplified treatment corresponds to no coupling of forces transverse to the 
direction of bacterial motion.  We 
hope that sensitive experiments can uncover the effects of these transverse
forces and hence yield more insight into polymerization-based force 
generation, though we expect
the effects to show up predominately in the geometrical prefactor or amplitude
of the curvature in Eqn.~\ref{EQN:rmsK} and not in the $1/\sqrt{n}$ dependence.
We expect our subsequent analysis, on distributions and autocorrelations
of the curvature, to be unaffected. 

The vectorial torque perpendicular to the 
bacterial direction of motion is Gaussian distributed, and the curvature is 
proportional to the {\em magnitude} of the torque. Within a bulk ($3d$)
geometry the curvature has distribution 
\begin{equation}
\label{EQN:intrinsic}
	P_I(x) = 2x e^{-x^2},
\end{equation}
where $x \equiv |K|/K_{rms}$ and $\int_0^\infty P_I(x)dx=1$.

\subsection*{Measuring curvature}

The curvature of the bacterial path, characterized by Eq.~\ref{EQN:rmsK} and
Eq.~\ref{EQN:intrinsic}, is {\em intrinsic} to a given random placement of 
$n$ filaments pushing against the bacterium.  This intrinsic
curvature is constant for fixed filament locations on the bacterial surface.
At any given time, the intrinsic curvature represents circular motion around 
the curvature axis. However, the instantaneous curvature axis is 
not necessarily parallel to the line of sight, so the apparent path would 
appear elliptical and have a non-uniform curvature.  We must also consider 
dynamical effects which change the direction of the curvature axis, both 
through rotation of the bacterium and through motion of the 
filaments on the bacterial surface. 

Most experimental work to date has been done with restricted geometries, 
such as the typical gap of several microns \cite{Kocks95} between a glass 
slide and its cover slip.  For thin enough samples, curvature out of the plane
will be restricted. If the axial angle $\phi$ measures the angle between the 
vectorial torque and the normal to the sample plane in the line of sight, 
then the apparent curvature as measured from a microscope or video image will be
\begin{equation}
K_{app} = |K| \cos{\phi}.
\end{equation}
Positive and negative curvatures correspond to clockwise and 
counterclockwise curved paths in the microscope image, respectively. 

A single bacterium with a fixed intrinsic curvature, $K$, 
would eventually uniformly explore $\phi \in [0,2 \pi]$ through 
rotational diffusion \cite{Berg93}.  As it does, the apparent curvature will
change. For a fixed intrinsic curvature, sampling at uniform time intervals, 
we would measure a distribution of apparent curvatures given by 
\begin{equation}
\label{EQN:apparent}
	P_A(x) = \frac{2}{\pi \sqrt{1-x^2}}, 
\end{equation}
where $x \equiv |K_{app}|/|K| \in [0,1]$. 

\begin{figure}
\centerline{\epsfxsize = 3.25truein
\epsfbox{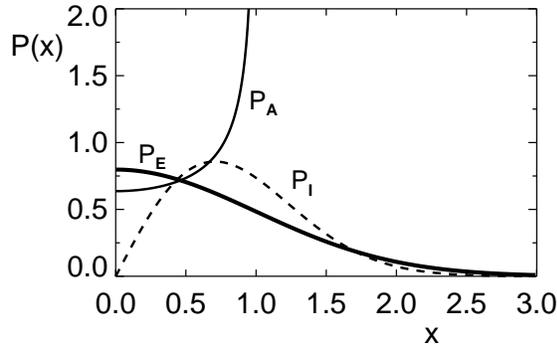}}
\caption{The curvature distributions $P_E(x)$ [thick solid], 
$P_A(x)$ [solid], and $P_I(x)$ [dashed] for ensemble, apparent, and 
intrinsic curvatures, respectively, vs the scaled curvature, $x$ (see text). 
We expect $P_E(x)$ if a single bacterium is tracked over large times or if 
a collection of bacteria are tracked, $P_A(x)$
if enough data can be gathered for a single bacterium 
before the intrinsic curvature changes, and $P_I(x)$ for a single 
bacterium if the curvature is extracted from a sufficiently long 
$3d$ trajectory in a bulk sample.
\label{FIG:distributions}}
\end{figure}

If an individual bacterium changes its intrinsic curvature in time, then
over sufficient time $K$ will explore the entire intrinsic
distribution, Eq.~\ref{EQN:intrinsic}. The ensemble distribution $P_E(x)$ 
of apparent curvatures will then be given by Eq.~\ref{EQN:apparent} 
convoluted with Eq.~\ref{EQN:intrinsic}:
\begin{eqnarray}
\label{EQN:ensemble}
P_E(x) &=& \int_x^\infty dy P_I(y) P_A(y/x)/y \nonumber \\ 
       &=& \sqrt{\frac{2}{\pi}} e^{-x^2/2},
\end{eqnarray}
where $x \equiv |K_{app}|/(K_{rms}/\sqrt{2})$, where the rms-apparent torque
is equal to $K_{rms}/\sqrt{2}$. $P_E(x)$ also follows directly
from the Gaussian distribution of each component of the torque. 
This ``ensemble'' distribution also characterizes 
the apparent curvatures of large groups of bacteria, 
since they will each have a different intrinsic curvature chosen
from Eq.~\ref{EQN:intrinsic} and each will have a random axial angle $\phi$. 
The differences between the ensemble, apparent, and intrinsic curvature
distributions is dramatic, as shown in Fig.~\ref{FIG:distributions}.

For bacteria in a bulk sample, two angles are needed to characterize the 
bacterial path with respect to the viewer --- a polar angle $\Psi$ 
with respect to the line of sight,
and an azimuthal angle $\phi_0$ about the line of sight. One must also 
specify the axial angle $\phi$ about the bacterial axis.
If we line up the bacterium to travel away from us
along the line of sight, before rotating it towards us by $\Psi$ in the 
$\phi_0$ azimuthal direction, and measure $\phi$ as the angle between 
the vectorial torque and $\phi_0$, then the apparent curvature is
\begin{equation}
K_{app,3d} = -|K| \cos{\phi}/\sin^2 \Psi.
\end{equation}
This is the curvature as measured from a microscope or video image, and 
takes values from $(-\infty,\infty)$ with no explicit $\phi_0$ dependence. 
Extremely large apparent curvatures 
are seen when bacteria are moving towards or away from the observer
with correspondingly small velocities, with $\Psi$ small. The 
apparent speed is $v_{app}= v \sin \Psi$, where $v$ is the 
bacterial speed along its path. A convenient quantity 
is obtained by ``normalizing'' the apparent 
curvature by multiplying by $v_{app}^2$, to obtain 
\begin{equation}
\label{EQN:knormalize}
	\tilde{K}_{app} \equiv K_{app,3d} v_{app}^2 = -\tilde{K} \cos \phi
\end{equation}
where $\tilde{K} \equiv |K v^2|$.  This also decreases the weight placed on 
apparently stationary bacteria, which can be hard to distinguish from other
objects, and simplifies the analysis. 
 $\tilde{K}_{app}$ will have the ensemble distribution $P_E(x)$ 
{\em if} $\phi$ is uniformly explored in
time.  Note that time or path-length weighted distributions will 
differ in bulk samples, since the apparent speed $v_{app}$ will vary 
dramatically even for a constant intrinsic bacterial speed.   The distributions
presented in this paper are for time-weighted sampling, appropriate for
video microscopy and/or for planar ($2d$) geometries. 

\section*{Curvature Dynamics}

For individual bacteria, the experimentally apparent 
variation of curvature from one moment to the next is striking 
\cite{Kocks95,Heinzen99,Gouin99}. Restricting ourselves to thin planar
samples, the angular diffusion of the bacterial orientation can lead to 
changing apparent curvatures through changing axial angles $\phi$ in
Eq.~\ref{EQN:knormalize}, and the {\em intrinsic}
curvature can also vary if filament locations on the bacterial surface
move significantly over time. We explore two cases, where active (pushing) 
filaments are removed and randomly replaced on the bacterial surface, and where 
active filament locations randomly diffuse. We will later apply
these cases to characterize filament repositioning 
for different motile systems. 

It is easiest to characterize changes in the net torque acting on the
bacterium with respect to a reference frame fixed to the bacterium. 
We consider the correlation of the intrinsic vector torques separated
in time by $\Delta t>0$:
\begin{eqnarray}
\label{EQN:torquecorr}
  A_N(\Delta t)	
		&\equiv& \langle \vec{N}(t) \cdot \vec{N}(t+\Delta t)\rangle 
			\nonumber \\
		&=& N_{RMS}^2  e^{-\Delta t/\tau}, 
\end{eqnarray}
where a static intrinsic curvature corresponds to $\tau=\infty$, and the 
average is over the initial time $t$. 
The second line follows directly if 
each filament has a lifetime $\tau$ after which it is replaced 
{\em randomly} on the rear of the bacterium by another filament. 
If new filaments are randomly
placed they will be uncorrelated with other filaments. The 
correlation will then be proportional to the fraction of filaments
that have not been replaced between the two times, i.e. $e^{-(t_2-t_1)/\tau}$.
Exponential decay also applies for actin filaments whose fast-growing 
barbed-end positions 
diffuse over the bacterial surface with diffusion constant $D$. 
Solutions of diffusion on a spherical surface of radius $b$ 
\cite{Koppel85} leads to $\tau = b^2/(2D)$. 
These results would apply directly to proteolysis/replacement or 
surface motion of bacterial proteins such as ActA or IcsA if active filament 
positions are localized to such bacterial surface features (see below). 
They should also apply to dynamics intrinsic to the actin tail through 
capping and nucleation of active actin filaments 
\cite{Machesky99b}, where capping is a loss mechanism. In this case, the 
timescale of of autocorrelation decay would depend on the details of filament
nucleation --- how new filaments are placed with respect to pre-existing
active filaments. 

Azimuthal diffusion will not affect the intrinsic curvature or its
correlations, but it can contribute to decay of correlations of the 
{\em apparent} curvature by changing the apparent curvature over time. 
Azimuthal diffusion obeys $\langle (\Delta \phi)^2 \rangle = 2 D_A \Delta t$, 
where $ \Delta \phi$ is the net angle of rotation in the time 
$\Delta t$ and $D_A$ is the diffusion constant. This has a direct effect on the 
correlation between apparent curvatures separated in time by $\Delta t>0$:
\begin{eqnarray}
\label{EQN:apparentcorr}
  A_{app}(\Delta t)	
		&\equiv& \langle K_{app}(t) 
					K_{app}(t+\Delta t) \rangle 
						\nonumber \\
		&=& \langle K_{rms}^2 \rangle e^{- \Delta t/\tau}
			\langle \cos \phi(t) \cos (\phi(t)+ \Delta \phi) 
				\rangle \nonumber \\
		&=& \langle K_{rms}^2 \rangle e^{- \Delta t/\tau}
			 e^{-D_A \Delta t}/2,
\end{eqnarray}
where $\tau$ is the decay time from intrinsic correlations in 
Eq.~\ref{EQN:torquecorr} and the average is over $t$ for a single bacterium. 
[We have used the identity 
$\langle e^{ix} \rangle = e^{-\langle x^2 \rangle/2}$ 
for Gaussian distributed $x$.] 
Of course, rotation of the bacterium around its long axis will lead to 
decaying autocorrelations only if active filament tips are localized to 
bacterial surface features. If not, azimuthal diffusion might not affect
the apparent curvature. 

For individual bacteria tracked for times much less
than $\tau$, the intrinsic curvature will appear constant. For
times much longer than $\tau$, each bacterium will sample the ensemble of 
intrinsic curvatures.  The characteristic timescale may be measured from the 
decay of curvature autocorrelations. Filament decay/replacement, 
filament diffusion, and axial rotation all contribute to 
exponential decay of the apparent curvature correlations. Their contributions 
to a particular motile system may be separated through independent measurements
or through systematic studies where parameters such as the particle size or
the cytoplasmic viscosity are varied. 

Random rotation and diffusion of bacterial positions will also contribute to 
the measured curvature and curvature autocorrelations. In principle this is a 
complicated hydrodynamic effect \cite{Pagonabarraga99} leading to exponential
asymptotic decay, however the timescales are very short 
($\tau \sim R^2 \rho/\eta \lesssim 10^{-5} s$ for a bacterium in a cell, 
where $R$ is the cell size and $\rho$ the cytoplasmic density)
compared to the measurement intervals in typical bacterial experiments 
(seconds).  The autocorrelation decay will effectively be discontinuous 
at $\Delta t =0$, where thermal and measurement jitter will contribute 
at $\Delta t=0$ but not for $\Delta t>0$.  To eliminate those contributions,
the experimental RMS curvature $K_{rms}$ 
should be extracted from the $\Delta t \rightarrow 0^+$ limit of the 
autocorrelations or should be fit from sufficiently long segments of the
bacterial path. 

Curvature autocorrelations in bulk samples are simple only when the intrinsic
curvature is extracted from full $3d$ tracking of the bacterial trajectory
(see e.g. \cite{Thar2000}). In that case, Eq.~\ref{EQN:torquecorr} will
describe the autocorrelation decay.
  
\section*{Motile systems}

In this section we discuss several specific motile systems, and use the 
details to refine our discussion of curved bacterial paths.  For illustrative
purposes, curvature has been estimated from published images of
{\em L. monocytogenes}.  This should be considered an 
order of magnitude estimate only.  We only analyze motion within 
{\em Xenopus laevis} cell extracts, since cellular organelles and cell 
membranes, which can locally affect bacterial trajectories through 
collisions and which are hard to control for in published images, are absent.  
Also absent in extracts is a polarized cytoskeleton, which could plausibly 
align bacterial motion in intact cells --- this could be explored through 
a systematic comparison of bacterial motion in cells and in cell extracts. 
It must be emphasized that {\em proper studies of curvature require 
unbiased data} and individual images previously selected for publication may be 
biased by aesthetic considerations. Distributions and autocorrelations 
require much more data than is available from published individual images, 
and will require analysis of video data. 

The details of the nanoscale mechanical connection between the actin filament 
tail and a particular bacterium or motile particle are not yet known, nor are
the details of the dynamics. Indeed, these details may differ for different
bacteria or for different natural or reconstituted cytoplasmic environments. 
We present some plausible scenarios below and indicate the expected results
of a curvature analysis in each.

\subsection*{\em Listeria monocytogenes}

Motile {\em L. monocytogenes} have a distribution of 
tail lengths ranging up to about $15$ $\mu m$ \cite{Tilney92},
and speeds of up to $0.4$ $\mu m/s$ \cite{Theriot92}. 
Mature {\em L. monocytogenes} are roughly cylindrical
Gram-positive bacteria, $1.5$ $\mu m$ long with a diameter of approximately 
$0.5 \mu m$ \cite{Gouin99}.  Polar surface expression of ActA is required 
\cite{Kocks92,Smith95} for motility. 
{\em L. monocytogenes} in {\em Xenopus} extract \cite{Theriot94} have apparent
curvatures of approximately $K \approx 0.08 \mu m^{-1}$, corresponding to 
$n=20$ filaments pushing on the bacterium. While this is a relatively small
number, it is consistent with electron microscopy images \cite{Gouin99}.

The role of ActA in polymerization is being uncovered \cite{Welch98} but
questions remain.  It is not yet clear whether individual filaments are 
associated with individual ActA molecules and if so, for how long.  
Direct mechanical attachment is possible and indeed indicated by optical 
tweezer studies \cite{Gerbal99}. Indirect attachment is also possible if 
the complex of ActA and cytoplasmic proteins serve as a source 
of monomeric actin with locally enhanced polymerization affinity. 
For example, profilin has been shown to interact with ActA through 
vasodilator-stimulated phosphoprotein (VASP) \cite{Smith95} and hence 
can provide a local "plume" of profilin-ATP-G-actin which in some conditions
can polymerize more readily than ATP-G-actin \cite{Pantaloni93}.
We can readily obtain the leading behavior of the 
steady-state concentration $C$ diffusing a distance $d$ from a 
disk-like source of radius $s$ and strength $C_0$ \cite{Crank75}, 
$C/C_0 = 2s/(\pi d)$. The diffusive plume would provide significantly enhanced 
polymerization only to distances on order the size $s$ of the ActA itself.
If ActA is well separated on the bacterial surface, then 
filament nucleation and growth would be closely associated with individual ActA 
molecules, which in turn would be held stationary by the external 
peptidoglycan layer of the bacterium.  The proteolysis induced 
lifetime $\tau$ of individual ActA is approximately $2$ hours {\em in vivo} 
\cite{Moors99}, and would directly contribute to the curvature 
autocorrelation decay given by Eq.~\ref{EQN:apparentcorr}.

If there is a connection between the bacterium and its tail \cite{Gerbal99},
then axial diffusion will be dramatically
reduced and autocorrelation decay due to proteolysis should dominate and 
could be directly observed. However, if 
the bacterium is not tightly attached to its tail, then it will rotate
with $D_A=k_B T/ C_A \approx 10^{-3}rad^2 s^{-1}$, corresponding to a 
timescale in Eq.~\ref{EQN:apparentcorr} of $1/D_A \approx 1000 s \ll \tau$,
where $C_A = (32 \pi/3) \eta (a^2-b^2)b^2/(2a-b^2S)$
is the viscous drag coefficient for axial rotation \cite{Perrin34}, 
and we take $\eta=30$ poise ($3 Pa\;s$) following \cite{Mogilner96}.

\subsection*{\em Shigella flexneri}

The Gram-negative {\em S. flexneri} has a similar motility mechanism to 
{\em L. monocytogenes} \cite{Bernardini89,Goldberg95}; for example, 
it  has a unipolar surface protein required for motility, IcsA. 
{\em S. flexneri} are 
about $2.3 \mu m$ long and $0.5 \mu m$ in diameter, and move at speeds 
comparable to {\em L. monocytogenes} \cite{Gouin99}. However, 
differences are observed between {\em S. flexneri} and {\em L. monocytogenes}.
The tails of {\em S. flexneri} appear to have fewer actin 
filaments than {\em L.  monocytogenes} \cite{Zeile96}.  IcsA is targeted 
to one bacterial pole in {\em S. flexneri} and may diffuse in the 
outer membrane \cite{Steinhauer99}, in contrast to 
{\em L. monocytogenes} where ActA is stationary. 

Curvature studies can help investigate these differences. For example, if 
fewer filaments are actually pushing the bacterium, rather than simply fewer
filaments involved in cross-linking, then according to Eq.~\ref{EQN:rmsK}, 
the curvature of the path of {\em S. flexneri} will be systematically larger. 
If actin filament tips are associated with individual IcsA, then 
diffusion of IcsA on the outer bacterial membrane
will contribute a term $e^{-\Delta t/\tau_{diff}}$ 
where $\tau_{diff}=b^2/(2D)$ in Eq.~\ref{EQN:apparentcorr}, in addition to 
the finite IcsA lifetime due to proteolysis \cite{Goldberg93}.
Unfortunately, {\em S. flexneri} are not motile in {\em Xenopus} extracts 
\cite{Kocks95,Gouin99}, but they do have qualitatively similar curvature to 
{\em L. monocytogenes} in intact cells \cite{Gouin99}.

\subsection*{Spotted-fever {\em Rickettsiae}}

The spotted-fever group of {\em Rickettsiae} use actin-based motility 
for intracellular movement \cite{Heinzen93}. While a surface protein 
(rOmpA) of motile {\em R. rickettsii} has been 
implicated in tail formation and has sequence similarity to a domain of IcsA 
\cite{Charles99}, its surface distribution and specific 
biochemical role have not yet been characterized.  The most studied species, 
{\em R. conorii} \cite{Gouin99} and {\em R. rickettsii} 
\cite{Heinzen99}, are roughly the same size as
 {\em L. monocytogenes}, but move only $1/3$ as fast.

While {\em Rickettsiae} are obligate pathogens that are not viable in 
{\em Xenopus} extracts, the curvature of their paths in intact cells 
is qualitatively smaller than that of {\em L. monocytogenes} or 
{\em S. flexneri} \cite{Gouin99}. Yet the tails in {\em R. conorii} are 
found to have very long parallel filaments, with relatively 
few filaments observed in cross-section electron-micrographs \cite{Gouin99}.  
It is worth considering two possibilities. The first
is that the filaments are not uniformly distributed on the bacterial surface. 
The numeric prefactor of the curvature Eq.~\ref{EQN:rmsK} will range from 
$0$ for polar filaments, to $\sqrt{3/2}$ for uniform distribution on a 
hemisphere (Eq.~\ref{EQN:distrib}), to $3/2$ if all filaments are along the 
outer edge at $r=b$.  A smaller than expected curvature can result from
clustering of active filaments near the pole. However this appears to be 
unlikely from the electron micrographs of decorated actin tails \cite{Gouin99}.
A second possibility is that the filaments are not randomly distributed on 
the bacterial surface. If the arrangement is symmetric, or regularly spaced, 
then the expected torque would be reduced. A paracrystalline surface structure 
is observed on {\em Rickettsiae} bacteria \cite{Palmer74}, though it is
unknown whether it affects polymerization dynamics.

\subsection*{Viruses and Vesicles}

Actin polymerization-based motility is exhibited by the 
{\em Vaccinia} virus, where $\sim 200 nm$ diameter oblate virus 
particles move at $0.05 \mu m/s$ with tails of length
$8 \mu m$ in HeLa cells \cite{Cudmore95}.  One of the interesting
unsolved puzzles of this system is that the virus always travels in the
symmetry direction, which is the orientation of highest drag. 

If filaments are localized to the viral surface by a specific protein
\cite{Frischknecht99}, and if it diffuses over the viral surface, 
the intrinsic curvature will change with
time with $\tau=b^2/(2 D)$ following Eq.~\ref{EQN:torquecorr} and 
\cite{Koppel85}, where $b$ is the vesicular radius and $D$ is the diffusion 
constant of the motility protein on the viral membrane.  Using $
D \approx 10^{-9} cm^2/s$, appropriate for diffusion within vesicular 
membranes, we have $\tau=0.05s$!  This is much less than the timescale of 
rotational diffusion and would dominate autocorrelation decay if present. 

Polymerization-based motility with curved comet-tails also occurs 
in motile vesicle systems, which are attractive systems for systematic study
since they have fluid outer layers, spherical geometry, and a variety of 
sizes.  Vesicle motion has been re-constituted in {\em Xenopus} egg extracts 
\cite{Ma98}, in endocytosed vesicles \cite{Merrifield99}, and in extracts of 
nematode sperm \cite{Italiano96}. In some systems the vesicle lipids directly
mediate actin polymerization \cite{Ma98,Rozelle2000}, in which case the 
filaments are unlikely to be localized to particular spots on the vesicle 
surface. Effective filament motion could still occur 
due to random nucleation and loss of filaments from the vesicle surface, and 
this is a possible mechanism of intrinsic curvature autocorrelation decay in 
bacterial systems as well. 

\subsection*{Microspheres}

A simplified {\em in vitro} system using small polystyrene 
microspheres, coated with purified ActA and 
added to {\em Xenopus lavis} egg extract, has been shown to 
reconstitute actin-based motility  \cite{Cameron99}.  Could 
different intrinsic curvatures be sampled by a single ``inert'' microsphere? 
The ActA has its transmembrane domain replaced by a $6 \times$ His repeat, and 
is non-specifically bound to the carboxylated polystyrene microsphere. 
It may be possible for the ActA to randomly  crawl on the microsphere 
surface without detaching (for example, see \cite{Jervis97}). 
Allowing a small surface diffusion rate, $D = 10^{-12} cm^2/s$, results 
in a decay time for correlations of $\tau  = b^2/(2D) \simeq 325 s$ for 
a $0.5 \mu m$ diameter microsphere.  Data from curvature studies may thus 
may be able to demonstrate diffusion of ActA on the microsphere. 

Microspheres are also good systems for systematic studies of various
radii $b$ with a constant surface density of ActA and, presumably, actin
filaments $\sigma$. In the autocorrelation decay, 
a finite filament lifetime would make a contribution that
scales as $\tau \sim {\rm const.}$, 
diffusion on the microsphere surface would have $\tau \sim b^2$, while
axial diffusion would have $1/D_A \sim b^3$ \cite{Perrin34}. Varying
the cytoplasmic viscosity, on the other hand, 
should only affect axial diffusion, with $1/D_A \sim \eta$. 

\section*{Summary}

Random filament interaction with the bacterial or particle surface
can explain the characteristic curved paths observed in 
polymerization-based motility systems, such as  {\em L. monocytogenes}, 
{\em S. flexneri}, spotted-fever {\em Rickettsiae}, 
{\em Vaccinia} virus, and motile lipid vesicles 
and microspheres. We distinguish between the intrinsic curvature, which can 
only be measured with the full three-dimensional trajectory of the bacterium
or particle, and the apparent curvature observed in microscope
images. We derived explicit distributions for these curvatures and showed how
they can uncover important qualitative differences between the various
polymerization-based motility 
systems. We showed, in Eq.~\ref{EQN:apparentcorr},  
how the lifetime and dynamics of 
surface-associated proteins, such as ActA or IcsA, affects the
evolution of the intrinsic curvature of the motion of individual bacteria, 
virus particles, vesicles, or protein-coated microspheres.   

Systematic experimental analysis of curvature in polymerization-based
motility systems has not yet
been done, but would supplement genetics, biochemistry, and 
microscopy by providing structural information about the interface between the 
actin tail and the bacterium. Curvature studies could estimate the number of 
filaments actively pushing the bacterium, the distribution
of these active filaments on the bacterial surface, their localization
with respect to motility protein complexes, and 
surface motility protein lifetime and diffusion on the bacterial surface. 
A similar analysis can be done for virus, vesicle, and microsphere systems. 

It is useful to summarize the specific applications of this 
analysis.  There are three.  {\em First}, the relation between curvature and 
number of filaments in the actin tail, Eqn.~\ref{EQN:rmsK}, can be compared with
electron-microscopy cross-sections and with normalized fluorescence studies. 
The relation between curvature and particle size, 
also in Eqn.~\ref{EQN:rmsK}, can be used for different sized particles with 
similar surface preparations, such as microspheres.  
{\em Second}, the distribution of observed curvatures 
is predicted to be Gaussian, Eqn.~\ref{EQN:ensemble}.  For particles
with constant intrinsic curvatures a qualitatively distinct distribution of 
apparent curvatures, Eqn.~\ref{EQN:apparent},
is expected.  If three-dimensional tracking of particles in a thick sample 
is used, then the intrinsic distribution would apply, Eqn.~\ref{EQN:intrinsic}.
{\em Third}, the variation of curvature in time is predicted to be described by 
an exponentially decaying autocorrelation function, Eqn.~\ref{EQN:apparentcorr}.
The timescale of autocorrelation decay, $\tau$, characterizes how the 
position of active filaments in the bacterial tail change. 
Studies of different size 
particles, or direct tracking of azimuthal particle rotation, could help to 
untangle the possible contributions to curvature autocorrelation decay.

We have made simplifying assumptions to facilitate our analysis.  
We have assumed particular particle shapes and surface 
distributions of filaments.  We also took the individual filament forces 
$f_0$ to be in the direction of particle travel. 
Different shapes, surface distributions, and filament orientations would 
change the numerical prefactor in Eqn.~\ref{EQN:rmsK}, though the  
curvature distributions would not be affected.  
We have also assumed that individual filaments are {\em independently} 
randomly located.  Non-random symmetric filament locations will 
result in curvatures much {\em less} than predicted in this paper. In contrast
filament distributions that not symmetric will lead to qualitatively larger
curvature than discussed here, where we assume a random but azimuthally 
symmetric distribution.  We have also assumed that the viscosity does
not vary strongly over bacterial length-scales. Strong viscous
heterogeneities, perhaps caused by particle motion itself, as well as 
local constraints posed by cellular organelles and membranes, will affect
particle trajectories. This could dominate the effects described here 
{\em in vivo}.

This work was supported financially by the Natural Sciences and
Engineering Research Council of Canada, {\it le Fonds pour la Formation de
Chercheurs et l'Aide \`a la Recherche du Qu\'ebec}, and by the 
Canadian Institute of Advanced Research Program in the Science of Soft
Surfaces and Interfaces. We would like to thank
Lisa Cameron, Julie Theriot, and Bob Heinzen for discussions, and 
Sarah Keller and Jennifer Robbins for critical readings of earlier versions
of the manuscript.


\end{document}